\documentclass[prb,twocolumn]{revtex4-1}

\usepackage{amsmath}
\usepackage{amsfonts}
\usepackage{graphicx}

\begin{document}

\title{Entanglement isn't just for spin}

\author{Daniel V. Schroeder}
\email{dschroeder@weber.edu}
\affiliation{Physics Department, Weber State University, Ogden, UT 84408-2508}

\date{\today}

\begin{abstract}

Quantum entanglement occurs not just in discrete systems such as spins, but also in the spatial wave functions of systems with more than one degree of freedom.  It is easy to introduce students to entangled wave functions at an early stage, in any course that discusses wave functions.  Doing so not only prepares students to learn about Bell's theorem and quantum information science, but can also provide a deeper understanding of the principles of quantum mechanics and help fight against some common misconceptions.  Here I introduce several pictorial examples of entangled wave functions that depend on just two spatial variables.  I also show how such wave functions can arise dynamically, and describe how to quantify their entanglement.

\end{abstract}

\maketitle

\section{Introduction}

Although Schr\"odinger introduced\cite{Schrodinger,SchrodingerCat} the term \textit{entanglement} to quantum mechanics in 1935,\cite{EPR} most physicists did not begin using the term until the 1990s or later.  Even today there are quantum mechanics textbooks in use that do not use the word ``entanglement'' at all.\cite{someTextbooks}  More importantly, our teaching often glosses over the underlying concept:  that for any quantum system with more than one degree of freedom, the vast majority of allowed states exhibit ``correlations'' or ``non-separability.''

When we finally introduce students to entangled states, it is usually in the context of spin systems, such as the singlet state of a pair of spin-1/2 particles.  This example is unparalleled for its mathematical simplicity and direct applicability to Bell's theorem and quantum information science.  However, spin systems are also rather abstract and disconnected from the spatial wave functions that are more familiar to most students.  Students also encounter entangled wave functions when they apply quantum mechanics to atoms, but there the concept of entanglement tends to get muddied by the complications of three spatial dimensions and identical particles.  Moreover, neither entangled spins nor entangled atomic wave functions are easy to visualize.

Fortunately, it is easy to include simple pictorial examples of entangled spatial wave functions in any course that discusses wave functions:  an upper-division quantum mechanics course, a sophomore-level modern physics course, and in many cases an introductory physics course.  The purpose of this paper is to illustrate some ways of doing so.

The following section introduces non-separable wave functions for a single particle in two dimensions.  Section~III then reinterprets these same functions for a system of two particles in one dimension.  Section IV explains how entanglement arises from interactions between particles, and Sec.~V shows how to quantify the degree of entanglement for a two-particle wave function.  Each of these sections ends with a few short exercises\cite{supplement} to help students develop a conceptual understanding of entanglement.  The appendix reviews some of the history of how the term ``entanglement'' finally came into widespread use, more than a half century after Schr\"odinger coined it.

\section{One particle in two dimensions}

\begin{figure}[b]
\centering
\includegraphics[width=159pt]{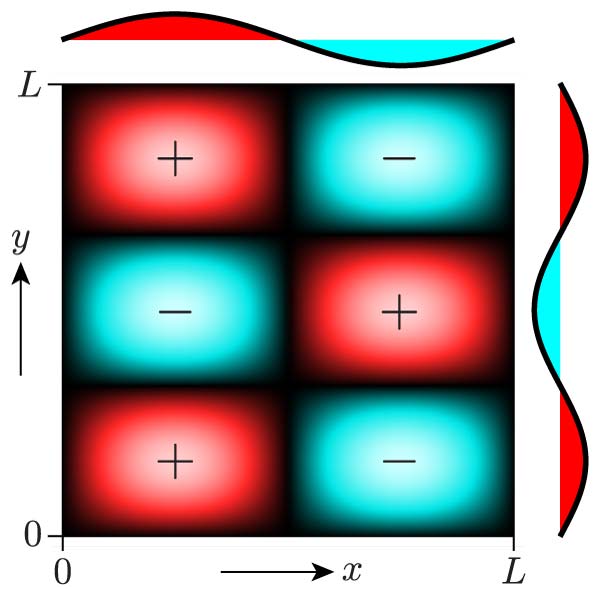}
\caption{Density plot of a typical separable solution to the time-independent Schr\"odinger equation for the two-dimensional square infinite square well.  This particular solution has $n_x=2$ and $n_y=3$.  Positive and negative portions of the wave function are indicated by the $+$ and $-$ symbols, and by color online; black represents a value of zero.  This wave function factors into a function of $x$ and a function of~$y$, drawn along the top and right.}
\label{Box23Plot}
\end{figure}

Imagine that you are teaching quantum mechanics to undergraduates and you have just finished covering wave mechanics in one dimension.  The natural next step is to explore wave mechanics in multiple dimensions, and a typical first example\cite{SquareWellTexts} is the two-dimensional square infinite square well, with potential energy
\begin{equation}
V(x,y) = \begin{cases}
0 & \text{when $0<x<L$ and $0<y<L$}, \\
\infty & \text{elsewhere}.
\end{cases}
\end{equation}
Inside this idealized potential well the separable solutions to the time-independent Schr\"odinger equation are
\begin{equation}
\psi(x,y) \propto \sin\Bigl(\frac{n_x \pi x}{L}\Bigr)\sin\Bigl(\frac{n_y\pi y}{L}\Bigr), \label{separablePsi}
\end{equation}
where $n_x$ and $n_y$ are positive integer quantum numbers.  The corresponding energies, assuming the particle is nonrelativistic, are proportional to $n_x^2 + n_y^2$.  Figure~\ref{Box23Plot} shows one of these wave functions.

After listing the allowed energies and perhaps drawing some of the separable wave functions, it is customary\cite{SquareWellTexts} to put this problem aside and go on to the next example---perhaps a three-dimensional infinite square well, or a central force problem.  Typically these further examples also admit separable solutions, and so our students run the risk of acquiring a serious misconception:

\begin{quote}
\textbf{Misconception 1:} All multidimensional wave functions are separable.
\end{quote}
Although I am not aware of any physics education research that documents the prevalence of this misconception, I think most of us who have taught quantum mechanics have encountered it.  Often students do understand that in order for the time-independent Schr\"odinger equation to have separable solutions the potential energy function must have a good deal of symmetry.  But as long as there is enough symmetry for separable solutions to exist, few students will spontaneously realize that these are not all the allowed wave functions.

Of course, mature physicists could never suffer from Misconception~1.  We know that quantum states live in a vector space, where every normalized linear combination of two or more allowed states is also an allowed state.  The separable wave functions of Eq.~(\ref{separablePsi}) are merely a set of basis states, from which all the others can be built.  But beginning students of quantum physics know none of this.  Many of them lack the vocabulary to even say it.

\begin{figure}[hbt]
\centering
\includegraphics[width=147pt]{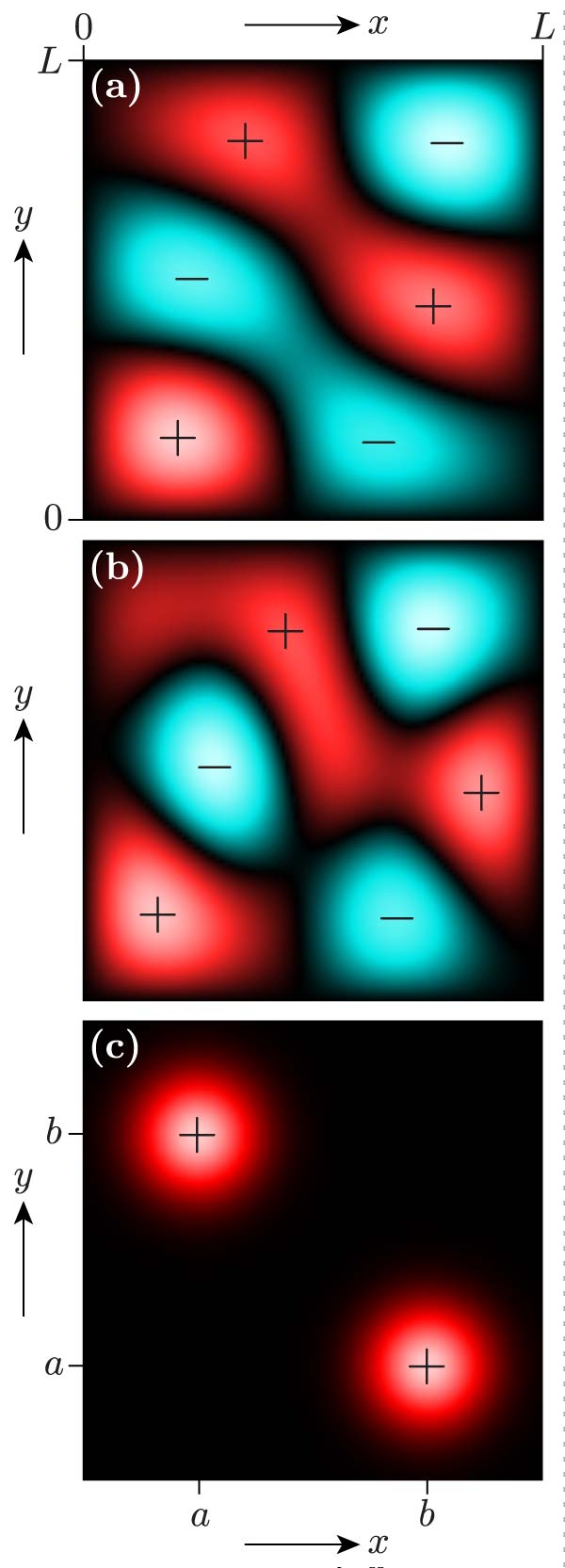}
\caption{Some non-separable wave functions for a particle in two dimensions. (a) A mixture of two degenerate square-well states as written in Eq.~(\ref{DegenerateMix}). (b) The same mixture as in (a), with a further admixture of the (5,1) state. (c) A ``cat state'' with isolated peaks at $(a,b)$ and $(b,a)$.}
\label{BoxNonSeparablePlots}
\end{figure}

Whether or not students know about vector spaces and orthonormal bases, it is easy enough to show them examples of superposition states.\cite{Park}  Figure~\ref{BoxNonSeparablePlots}(a) shows the state
\begin{equation}
\psi \propto \sin\Bigl(\frac{2\pi x}{L}\Bigr) \sin\Bigl(\frac{3\pi y}{L}\Bigr)
 + \frac12 \sin\Bigl(\frac{3\pi x}{L}\Bigr) \sin\Bigl(\frac{2\pi y}{L}\Bigr),
 \label{DegenerateMix}
\end{equation}
in which I have combined an admixture of the degenerate $(3,2)$ state with the $(2,3)$ state of Fig.~\ref{Box23Plot}.  Although it is built out of two separable pieces, this function is not itself separable: you cannot factor it into a function of~$x$ times a function of~$y$.  You can readily see from the plot that its $x$ dependence changes as you vary~$y$, and vice-versa.

Separability, or the lack thereof, is not merely a mathematical abstraction.
A separable wave function has the important \textit{physical} property that a measurement of one degree of freedom has no effect on a subsequent measurement of the other degree of freedom.  For example, if a particle is in the state shown in Fig.~\ref{Box23Plot} and you measure its $x$ coordinate and happen to obtain $L/4$, the probability distribution for a subsequent measurement of its $y$ coordinate is still proportional to $\sin^2(3\pi y/L)$, exactly the same as before you measured~$x$.  On the other hand, if the particle starts out in the state shown in Fig.~\ref{BoxNonSeparablePlots}(a) and you measure its $x$ coordinate and happen to obtain $L/4$, a subsequent measurement of $y$ is then considerably more likely to yield values near $L/4$, and less likely to yield values near $3L/4$, than it was before you measured~$x$.  In this case the outcomes of the two measurements are correlated; we could even say they are \textit{entangled}---although that word is usually reserved for systems of two or more distinct particles, as discussed in the following section.

\begin{figure}[t]
\centering
\includegraphics{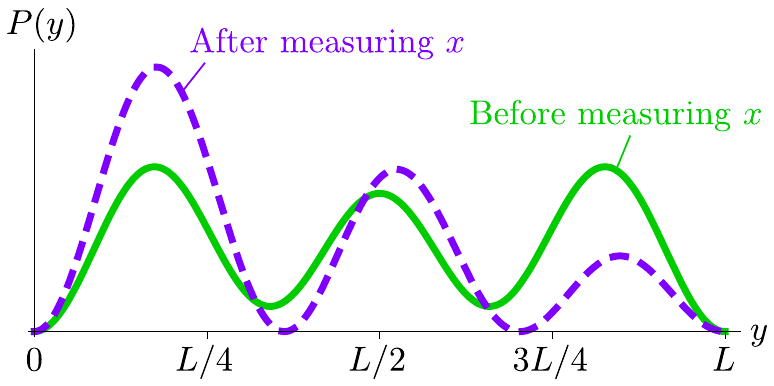}
\caption{Probability distributions for a measurement of $y$ on the state shown in Fig.~\ref{BoxNonSeparablePlots}(a), before measuring~$x$ (solid) and after measuring $x$ and obtaining the result $L/4$ (dashed).}
\label{BeforeAndAfterDistributions}
\end{figure}

Although the probability claims made in the previous paragraph should be fairly intuitive just from looking at the wave function density plots, they can also be quantified.  If you measure $y$ before~$x$, then you calculate the probability distribution for~$y$ by integrating the square of the wave function over all~$x$:
\begin{equation}
P(y) = \int_0^L \!|\psi(x,y)|^2\,dx \quad\text{(before measuring~$x$)}.
\label{yProbDensity}
\end{equation}
On the other hand, if you measure $x$ first and obtain the result $x_0$, then you calculate the probability distribution for a subsequent measurement of~$y$ by setting $x=x_0$ in the wave function (to ``collapse'' it along the measurement direction), renormalizing it, and then squaring:
\begin{equation}
P(y) \propto |\psi(x_0,y)|^2 \quad\text{(after measuring~$x$)}.
\end{equation}
Figure~\ref{BeforeAndAfterDistributions} compares these two probability distributions for the wave function shown in Fig.~\ref{BoxNonSeparablePlots}(a) and $x_0=L/4$.

In constructing the superposition state shown in Fig.~\ref{BoxNonSeparablePlots}(a) I chose to mix two basis states with the same energy, and therefore the result is still a solution to the time-independent Schr\"odinger equation.  But this was a pedagogically poor choice on my part, because it could reinforce another common misconception:

\begin{quote}
\textbf{Misconception 2:}  All allowed wave functions must satisfy the time-independent Schr\"odinger equation.
\end{quote}
Physics education researchers have convincingly documented\cite{ZhuSingh} the prevalence of this misconception, but even without documentation it should not come as a surprise, because we expect  students of quantum mechanics to spend so much of their time solving the time-independent Schr\"odinger equation.\cite{Styer1996}

A better example of a superposition state might therefore be the one shown in Fig.~\ref{BoxNonSeparablePlots}(b), which adds a component of the higher-energy $n_x=5$, $n_y=1$ state to the superposition of Fig.~\ref{BoxNonSeparablePlots}(a) and Eq.~(\ref{DegenerateMix}).  Again, this wave function is non-separable and therefore has the property that a measurement of $x$ will change the probability distribution for a subsequent measurement of~$y$ (and vice-versa).

But why restrict our attention to superpositions of two or three square-well basis states?  Figure \ref{BoxNonSeparablePlots}(c) shows an even clearer example of non-separability:  a ``cat state''\cite{CatState} consisting of two isolated peaks, one centered at coordinates $(a,b)$ and the other at $(b,a)$.  Of course the completeness of the basis states guarantees that this state can be expressed as a linear combination of them, but if the goal is to understand non-separability (or ``entanglement'' of $x$ and~$y$), then there is no need to mention any basis or even to assume that this particle is inside an infinite square well.  By inspection we can see that if a particle is in this cat state then a measurement of either $x$ or $y$ will have a 50-50 chance of giving a result near either $a$ or $b$.  However, if we measure $x$ first and happen to get a result near~$b$, then a subsequent measurement of $y$ is guaranteed to give a result near~$a$.

\begin{figure}[t]
\centering
\includegraphics[width=213pt]{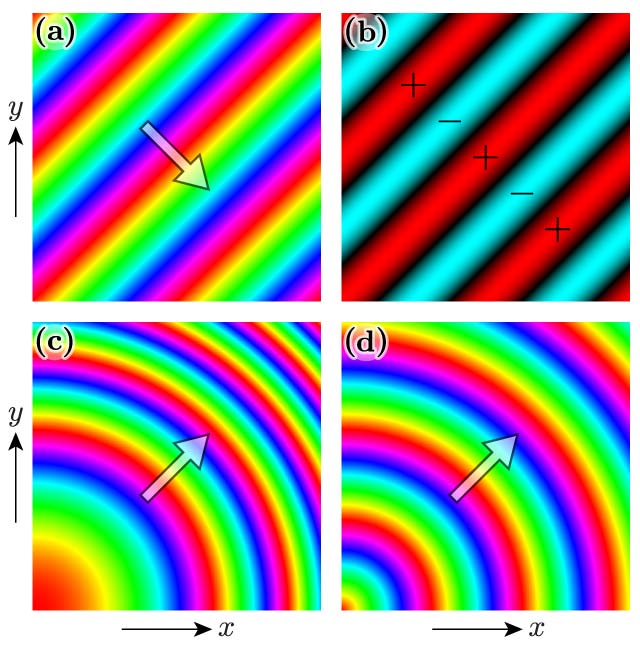}
\caption{Four two-dimensional wave functions built from complex exponentials: (a) a separable linear wave, $\exp[{ik(x-y)}]$; (b) a non-separable superposition of opposite-moving linear waves, $\exp[{ik(x-y)}]+\exp[{-ik(x-y)}]$; (c) a separable circular wave, $\exp[{ik^2(x^2+y^2)}]$; (d) a circular wave that is separable in polar coordinates but not in rectangular coordinates, $\exp[{ik\sqrt{x^2+y^2}}]$.  The color hues (online) indicate the complex phases, with the arrows pointing in the direction of increasing phase.}
\label{ComplexFunctions}
\end{figure}

Further examples abound.  For instance, we can consider complex-valued wave functions such as those shown in Fig.~\ref{ComplexFunctions}.  Each of these plots uses color hues to represent the complex phases,\cite{VisualQM} and shows only a square portion of a function that extends over a larger area.  Recognizing separable and non-separable functions from such plots can be tricky, because the phase factor $e^{i\theta}$ shifts the hues rather than scaling the brightness.  

\begin{quote}
\textbf{Exercise 1:} Determine the missing normalization constants in Eqs.\ (\ref{separablePsi}) and (\ref{DegenerateMix}).
\end{quote}

\begin{quote}
\textbf{Exercise 2:} Use a computer to reproduce Fig.~\ref{BoxNonSeparablePlots}(b), adjusting the relative coefficient of the (5,1) state to obtain a good match.  (A tutorial on plotting wave functions with Mathematica is included in the electronic supplement to this paper.\cite{supplement})  Then find the overall normalization constant and write down the full formula for this wave function.  Calculate and plot the probability distribution $P(y)$ for a measurement of $y$ for this state, both before any measurement of~$x$ is performed and after measuring~$x$ and obtaining the value~$3L/4$.
\end{quote}

\begin{quote}
\textbf{Exercise 3:} Write down a qualitatively accurate formula, in terms of Gaussian functions, to represent the ``cat state'' shown in Fig.~\ref{BoxNonSeparablePlots}(c).  Show both pictorially and algebraically that this function \textit{is} separable if you rotate the coordinate axes by 45$^\circ$.  Thus, the ``entanglement'' of a two-variable wave function can be a coordinate-dependent concept.  Describe (and draw) at least two conceptually distinct ways in which you could modify this wave function so that it is not separable in any rotated coordinate system.
\end{quote}

\begin{quote}
\textbf{Exercise 4:} Suppose that you measure the components $p_x$ and $p_y$ of the momentum for a particle with the wave function shown in Fig.~\ref{ComplexFunctions}(a).  What values might you obtain (in terms of~$k$), with what probabilities?  Answer the same question for the wave function shown in Fig.~\ref{ComplexFunctions}(b).  Are the outcomes of the $p_x$ and $p_y$ measurements correlated?  Explain.
\end{quote}

\begin{quote}
\textbf{Exercise 5:} For the wave function shown in Fig.~\ref{ComplexFunctions}(c), suppose that you measure the $y$ component of the momentum and obtain a value near zero.  What does this tell you about the particle's location?  What does it tell you about the $x$ component of the particle's momentum?  Has the probability distribution for the $x$ component of the momentum changed as a result of the measurement?  Answer the same questions for the wave function shown in Fig.~\ref{ComplexFunctions}(d).
\end{quote}

\section{Two particles in one dimension}

In standard usage, the word \textit{entanglement} seems to be reserved for correlations between two particles, rather than between two different degrees of freedom ($x$ and $y$ in the previous section) of a single particle.  But this restriction is an arbitrary matter of semantics, because every wave function described in the previous section can be reinterpreted as a wave function for a system of \textit{two} particles in \textit{one} dimension, merely by relabeling $(x,y)\rightarrow(x_1,x_2)$.

Before proceeding to discuss a system of two particles, however, we need to confront a third misconception:

\begin{quote}
\textbf{Misconception 3:}  Every particle has its own wave function.
\end{quote}
Again I am not aware of any research to document the prevalence of this misconception.\cite{StyerAgain}  But I routinely see a look of surprise on students' faces when they learn otherwise, and I have even encountered this misconception among PhD physicists.  We reinforce it whenever we (and our chemistry colleagues) teach atomic physics and speak of the first two electrons being in the 1s shell, the next two in the 2s shell, and so on.  The English language naturally evokes classical images of particular objects in particular places---not entangled quantum states.

The best way to fight Misconception~3 is to give students plenty of opportunities to work with entangled two-particle wave functions:  plot them, interpret them in words, and do calculations with them.  Even for those who accept in the abstract that a two-particle system has only a single wave function that cannot in general be factored, working with specific examples can deepen understanding and build intuition.  Note that each point on a density plot of the two-dimensional wave function now gives the joint amplitude for finding particle~1 at $x_1$ \textit{and} particle~2 at~$x_2$.  To find the probability density for a position measurement of just one particle, we must integrate $|\psi|^2$ over the position of the other particle as in Eq.~(\ref{yProbDensity}).  Mentally switching between one-dimensional physical space and two-dimensional configuration space requires a good deal of practice.\cite{McIntyre}

With the replacement $(x,y)\rightarrow(x_1,x_2)$, the two-dimensional square infinite square well becomes a system of two particles confined in a one-dimensional infinite square well.  Alternatively, the two particles could be confined in two separate one-dimensional wells.  To make a precise analogy we must assume that the two particles are distinguishable, either by being in separate potential wells or by some other physical property; otherwise there would be symmetry constraints on the two-particle wave function.  The separable wave functions of Eq.~(\ref{separablePsi}) are still energy eigenfunctions as long as the particles do not interact, and the energy eigenvalues are then the same as before if the particles have equal masses.  For the two-particle system, however, it is more natural to think about separate energy measurements for the two degrees of freedom.  For example, if the system is in the state depicted in Fig.~\ref{Box23Plot}, then we know that particle~1 has four units of energy and particle~2 has nine units, relative to the ground-state energy for a single particle.

Whether or not the two particles are confined inside an infinite square well, it is easy to construct two-particle wave functions that are entangled, that is, not separable.  We can form combinations of two or three of the separable square-well basis states, as shown in Figs.~\ref{BoxNonSeparablePlots}(a) and~(b).  We can imagine ``cat states'' with two or more separated peaks, as in Fig.~\ref{BoxNonSeparablePlots}(c).  And we can build states out of complex exponential functions, as shown in Fig.~\ref{ComplexFunctions}.  The exercises below explore all of these types of entangled states.

\begin{quote}
\textbf{Exercise 6:}  For the wave function shown in Fig.~\ref{BoxNonSeparablePlots}(a), with $(x,y)\rightarrow(x_1,x_2)$, what are the possible outcomes, and their probabilities, of a measurement of the energy of particle~2?  Suppose next that you first measure the energy of particle~1, and find that it has four units of energy (in terms of the single-particle ground-state energy); now what can you predict about the energy of particle~2?  What if instead you had found that particle~1 has nine units of energy?
\end{quote}

\begin{quote}
\textbf{Exercise 7:}  Repeat the previous problem for the wave function shown in Fig.~\ref{BoxNonSeparablePlots}(b).  Consider all possible outcomes of the measurement of the energy of particle~1.  (Before working this exercise you should work Exercise~2.)
\end{quote}

\begin{quote}
\textbf{Exercise 8:}  When we reinterpret the ``cat state'' of Fig.~\ref{BoxNonSeparablePlots}(c) to apply to two particles in one dimension, it is tempting to assume that each of the two wave function peaks represents one of the two particles.  Why is this assumption wrong?  What \textit{does} each of the peaks represent?  Explain carefully.
\end{quote}

\begin{quote}
\textbf{Exercise 9:}  For the ``cat state'' shown in Fig.~\ref{BoxNonSeparablePlots}(c), with $(x,y)\rightarrow(x_1,x_2)$, sketch the probability distributions $P(x_1)$ and $P(x_2)$ for measurements of the positions of the two particles.  Now sketch at least two other wave functions, one entangled and one not, that are different from the one shown yet still yield the same probability distributions for both particles.  Explain the physical differences among all three wave functions, in terms of outcomes of successive measurements of $x_1$ and~$x_2$.  For instance, if you measure $x_1$ first and obtain a value near~$a$, what can you predict about the outcome of a subsequent measurement of~$x_2$?
\end{quote}

\begin{quote}
\textbf{Exercise 10:}  For the wave function shown in Fig.~\ref{ComplexFunctions}(b), with $(x,y)\rightarrow(x_1,x_2)$, sketch the probability distributions $P(p_1)$ and $P(p_2)$ for measurements of the momenta of the two particles.  Suppose now that you measure $p_1$ and obtain a positive value; what can you now predict about the outcome of a subsequent measurement of~$p_2$?
\end{quote}

\begin{quote}
\textbf{Exercise 11:}  Imagine an infinite square well containing two particles whose wave function has the form $\sin(\pi x_1/L)\sin(\pi x_2/L)$ (to meet the boundary conditions) times a Gaussian factor that tends to put the two particles close to each other:  $\exp[-((x_1{-}x_2)/a)^2]$, where $a=L/10$.  Sketch this wave function, then sketch the probability distribution for a measurement of~$x_2$.  Now imagine that you measure $x_1$ and happen to obtain the result $0.3L$; sketch the new probability distribution for a subsequent measurement of~$x_2$.  (Instead of merely sketching, you could use a computer to make quantitatively accurate plots.)
\end{quote}

\begin{quote}
\textbf{Exercise 12:}  Suppose that for a calculation or a plot you need to know the wave function of a particle in one dimension, confined within a width~$L$, to a resolution of $L/100$.  For a single particle you then need to know 100 complex numbers (minus one if we neglect the normalization constant and unphysical overall phase).  How many numbers must you know to represent the wave function of two particles at this same resolution?  Three particles?  If your computer has eight gigabytes of memory and each complex number takes up eight bytes, what is the maximum number of particles for which your computer can store an arbitrary wave function?\cite{StyerCourseNotes}
\end{quote}

\section{Dynamics}

Just because a quantum state is allowed doesn't mean it will occur in practice.  Students will naturally wonder how to create entangled two-particle states in the real world.  We owe them an answer, even if we illustrate that answer with idealized examples in one spatial dimension.

The answer, in a word, is \textit{interactions}:  particles tend to become entangled when they interact with each other.\cite{IdenticalParticles}  Schr\"odinger himself said it well:\cite{Schrodinger}

\begin{quote}
When two systems, of which we know the states by their respective representatives [i.e., wave functions], enter into temporary physical interaction due to known forces between them, and when after a time of mutual influence the systems separate again, then they can no longer be described in the same way as before, viz.\ by endowing each of them with a representative of its own.  I would not call that \textit{one} but rather \textit{the} characteristic trait of quantum mechanics, the one that enforces its entire departure from classical lines of thought.  By the interaction the two representatives (or $\psi$-functions) have become entangled.
\end{quote}

\begin{figure}[t]
\centering
\includegraphics[width=142pt]{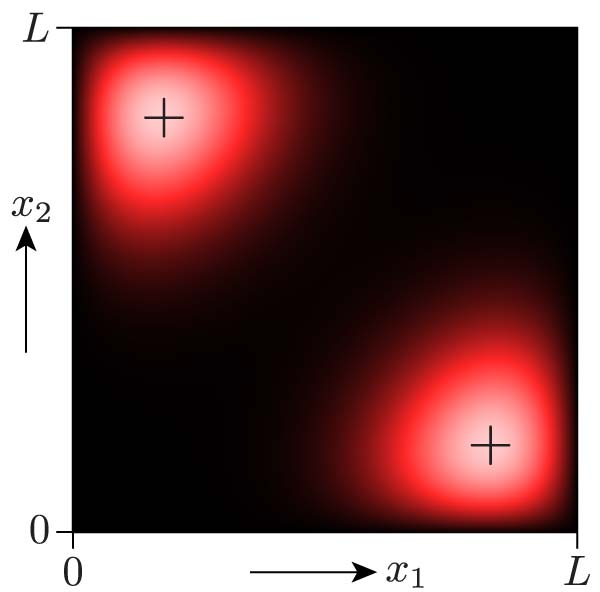}
\caption{Entangled ground-state wave function for a system of two equal-mass but distinguishable particles confined to a one-dimensional infinite square well and interacting via the repulsive potential of Eq.~(\ref{GaussianRepulsion}).  In natural units with $\hbar$, the particle mass, and the well width all equal to~1, the parameters of the potential are $V_0=200$ and $a = 0.5$.  See Ref.~\onlinecite{VariationalRelaxation} for details on how this wave function was calculated.  Software for doing such calculations is included in the electronic supplement to this paper.\cite{supplement}}
\label{TwoInteractingParticlesGroundState}
\end{figure}

For a specific example, let us first go back to the familiar context of two equal-mass (but distinguishable) particles trapped in a one-dimensional infinite square well.  If we merely add an interaction term of the form $V(x_2-x_1)$ to the Hamiltonian of this system, then all the stationary-state wave functions will be entangled.\cite{TwoInteractingInBox}  For example, Fig.~\ref{TwoInteractingParticlesGroundState} shows the ground-state wave function for the case of a repulsive Gaussian interparticle interaction,
\begin{equation}
V(x_1,x_2) = V_0 e^{-(x_2-x_1)^2/a^2}. \label{GaussianRepulsion}
\end{equation}
In the two-dimensional configuration space of this system, this potential is simply a barrier running along the main diagonal, centered on the line $x_2=x_1$.  The barrier divides the square region into a double well, so the system's ground state consists of a symmetrical double peak, similar to the cat state of Fig.~\ref{BoxNonSeparablePlots}(c).  In other words, as we would expect, the repulsive interaction tends to push the particles to opposite sides of the one-dimensional square well, but neither particle has a preference for one side or the other.

\begin{figure}[t]
\centering
\includegraphics[width=141pt]{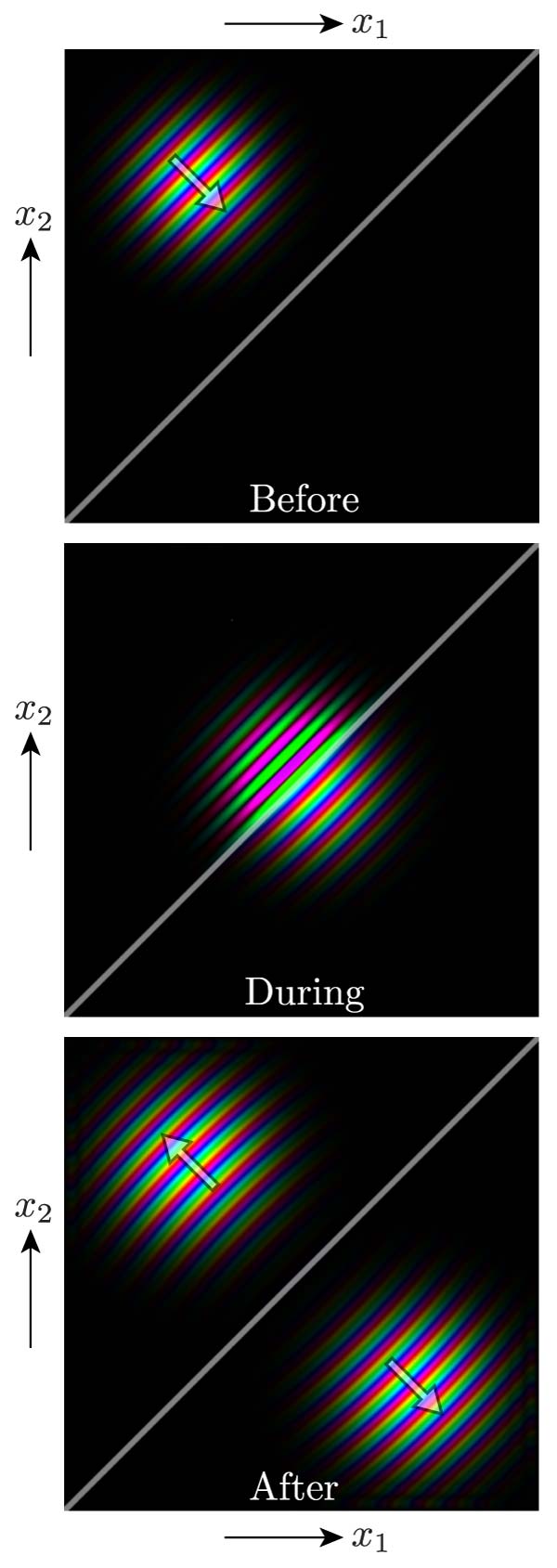}
\caption{Sequence of three frames showing a two-particle scattering event, in one spatial dimension, as calculated by a numerical integration of the time-dependent Schr\"odinger equation.  The particles interact via a short-range repulsive rectangular barrier, Eq.~(\ref{RectangularBarrier}), indicated by the gray diagonal band.  The brightness indicates the  magnitude of the wave function (scaled differently in each frame), while the color hues (online) indicate the phase.  The arrows show the direction of increasing phase, i.e., the direction of motion.  Software for performing simulations of this type is provided in the electronic supplement to this paper.\cite{supplement}}
\label{ScatteringSequence}
\end{figure}

Figure \ref{ScatteringSequence} shows an example involving a temporary interaction of the type that Schr\"odinger described.  Here two equal-mass particles, in one dimension, are initially in a state consisting of separated Gaussian wave packets, moving toward each other.  (Note that these physically separated wave packets appear as a single peak in two-dimensional configuration space.)  The particles interact via a short-range finite rectangular barrier,
\begin{equation}
V(x_1,x_2) = \begin{cases}
V_0 & \text{for $|x_2-x_1| < a$}, \\
0 & \text{otherwise},
\end{cases}
\label{RectangularBarrier}
\end{equation}
where the parameters $V_0$ and $a$ have been chosen to give transmission and reflection probabilities that are approximately equal.  After the interaction, therefore, the particles are in an entangled state whose probability distribution resembles that of the cat state in Fig.~\ref{BoxNonSeparablePlots}(c), but with peaks moving away from each other as time goes on.  One peak puts the two particles back near their starting positions, indicating reflection; the other peak puts them in interchanged locations, indicating transmission or tunneling.  Notice that for this state, a measurement of one particle's position affects not only the probability distribution for the other particle's position, but also the probability distribution for its momentum.\cite{EntanglementInScattering, CoordinatesRemark}

Fundamental though they are, examples like these rarely appear in quantum mechanics textbooks.  The reason is probably that despite their conceptual simplicity, a quantitative treatment of either scenario requires numerical methods.  The wave function plotted in Fig.~\ref{TwoInteractingParticlesGroundState} was calculated using a variational-relaxation algorithm,\cite{supplement, VariationalRelaxation} while Fig.~\ref{ScatteringSequence} is the result of a numerical integration of the time-dependent Schr\"odinger equation.\cite{supplement, Maestri}  Although neither calculation takes more than a few seconds on today's personal computers, learning to do such calculations is not a standard part of the undergraduate physics curriculum.  Teaching students to do these numerical calculations would serve the dual purpose of augmenting their computational skills and helping them develop intuition for entangled two-particle systems.  On the other hand, students who have already studied single-particle examples of double-well bound states and wave packet scattering should not need any computational skills to make qualitative predictions or to understand the pictorial results.  The bottom line is that interactions between particles generically create entanglement.

\begin{quote}
\textbf{Exercise 13:}  Describe and sketch the first excited state for the system whose ground state is depicted in Fig.~\ref{TwoInteractingParticlesGroundState}.  (Hint: What does the first excited state look like for a one-dimensional double well?)
\end{quote}

\begin{quote}
\textbf{Exercise 14:}  Consider the system of two particles in an infinite square well, with a Gaussian interaction as in Eq.~(\ref{GaussianRepulsion}), but with $V_0<0$, so the interaction is attractive.  Sketch what the ground-state wave function of this system might look like, and interpret it physically in terms of measurements of the two particles' positions.
\end{quote}

\begin{quote}
\textbf{Exercise 15:}  When two particles do \textit{not} interact with each other, the system's potential energy has the form $V(x_1,x_2) = V_1(x_1) + V_2(x_2)$.  Prove that in this case (a) the time-independent Schr\"odinger equation separates into an equation for each particle, so that there exists a complete set of unentangled stationary states; and (b) if the system's initial state is not entangled, then its state will remain unentangled as time passes.
\end{quote}

\section{Quantifying entanglement}

The scattering example of the previous section makes it clear that not all interactions produce equal amounts of entanglement.  By adjusting the range and strength of the interaction potential $V(x_1-x_2)$, we can obtain transmission probabilities ranging from 0 to~1, and in either of these limits the final state would not be entangled.  There seems to be a sense in which the entanglement is maximized when the reflection and transmission probabilities are equal.

More generally, consider any wave function built as a normalized superposition of two separable and orthogonal terms:
\begin{equation}
\psi(x_1,x_2) = \alpha f_1(x_1)g_1(x_2) + \beta f_2(x_1)g_2(x_2), \label{TwoTermSuperposition}
\end{equation}
where $f_1$ and $f_2$ are orthonormal functions of $x_1$, $g_1$ and $g_2$ are orthonormal functions of $x_2$, and $|\alpha|^2+|\beta|^2=1$.  There is no entanglement when $\alpha$ or $\beta$ is zero, and we intuitively expect the ``amount'' of entanglement to increase as $|\alpha|$ and $|\beta|$ approach each other, reaching a maximum when $|\alpha|^2=|\beta|^2=1/2$.  But how can we quantify this intuition to obtain a formula for the amount of entanglement?

A general approach\cite{Grobe} is to calculate a quantity called the \textit{interparticle purity} of the two-particle state:
\begin{align}
p_{12} &= \int\!\!\int\!\!\int\!\!\int\psi(x_1,x_2)\psi^*(x_1^\prime,x_2) \nonumber \\
&\qquad\qquad\quad\cdot\psi(x_1^\prime,x_2^\prime)\psi^*(x_1,x_2^\prime)\,dx_1 dx_2 dx_1^\prime dx_2^\prime. \label{purity}
\end{align}
Experts may recognize this quantity as the trace of the squared one-particle reduced density matrix; for the rest of us, the best way to develop an understanding of this quantity is to work out some special cases.

First notice that if $\psi(x_1,x_2)$ is separable, then each of the four integrals in Eq.~(\ref{purity}) becomes a simple normalization integral, so $p_{12}=1$.

Next consider the two-term superposition of Eq.~(\ref{TwoTermSuperposition}).  Plugging this expression into Eq.~(\ref{purity}) results in 16 terms, but 14 of them are zero by orthogonality and the other two reduce to normalization integrals, yielding the simple result
\begin{equation}
p_{12} = |\alpha|^4 + |\beta|^4 = \textstyle 2(|\alpha|^2-\frac12)^2 + \frac12,
\label{TwoTermPurity}
\end{equation}
which equals 1 when $\alpha$ or $\beta$ is zero and reaches a minimum of 1/2 when $|\alpha|^2=|\beta|^2=1/2$.  Thus the interparticle purity is inversely related to the intuitive notion of entanglement described above, at least for a wave function of this form.

The following exercises explore the interparticle purity through further examples, and the software in the electronic supplement\cite{supplement} calculates $p_{12}$ for scenarios of the types considered in the previous section.  In general, the lower the value of $p_{12}$, the more a measurement on one particle tends to change the probability distribution for a subsequent measurement on the other particle.

\begin{quote}
\textbf{Exercise 16:}  Write out a full derivation of Eq.~(\ref{TwoTermPurity}), showing which terms are nonzero, which are zero, and why.
\end{quote}

\begin{quote}
\textbf{Exercise 17:}  Work out the formula for the interparticle purity of a superposition state of the form of Eq.~(\ref{TwoTermSuperposition}), but with three terms, still built from orthogonal functions, instead of just two.  What is the smallest possible value of~$p_{12}$ for such a state?  Can you generalize to a superposition of four or more such terms?
\end{quote}

\begin{quote}
\textbf{Exercise 18:}  Determine $p_{12}$ for each of the three wave functions depicted in Fig.~\ref{BoxNonSeparablePlots}, reinterpreted for two particles in one dimension.  Before doing this for (b) you should work Exercises 2 and~17.
\end{quote}

\begin{quote}
\textbf{Exercise 19:}  Make a rough estimate of the interparticle purity of the wave function considered in Exercise~11, representing two particles in an infinite square well that tend to be much closer together than the size of the well.  What happens in the limiting cases $a\rightarrow\infty$ and $a\rightarrow0$?  (You may also wish to calculate $p_{12}$ numerically.  To do so, it is probably best to sample the wave function on a grid to make a matrix $\Psi$ of values; then you can show that $p_{12}$ is proportional to the trace of $(\Psi^\dagger\Psi)^2$, or simply the trace of $\Psi^4$ if $\Psi$ is real and symmetric.)
\end{quote}

\begin{quote}
\textbf{Exercise 20:}  Equation (\ref{purity}) generalizes straightforwardly to systems in more than one spatial dimension.  Consider, then, an ordinary hydrogen atom, consisting of an electron and a proton, in its ground state.  The average distance between the two particles is then known to be on the order of $10^{-10}$~m.  If the atom as a whole is in a state that is spread over a volume of a cubic millimeter, what is the approximate interparticle purity of the two-particle state?
\end{quote}

\section{Discussion}

The goal of this paper is to illustrate some ways of introducing students to entangled wave functions at a relatively early stage in their physics education.  There are at least three reasons to do so.

First, as emphasized above, we want to prevent misconceptions.  When students learn only about separable wave functions they can develop an over-simplified view of how quantum mechanics works.

\begin{figure}[t]
\centering
\includegraphics{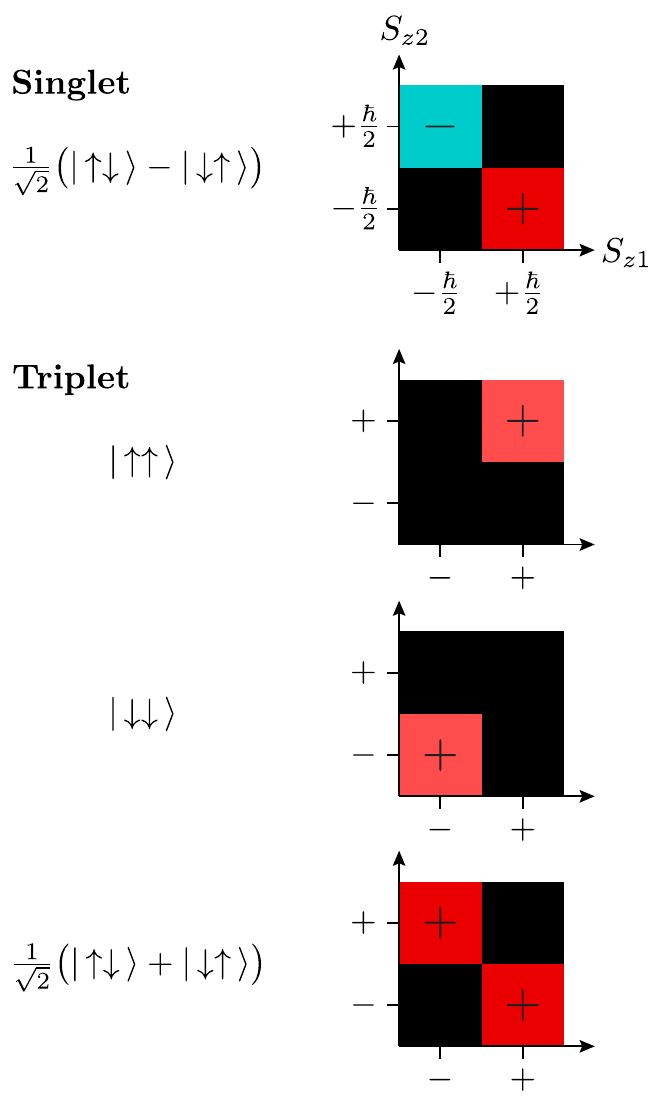}
\caption{Examples of $2\times2$ ``density'' plots for the singlet and triplet states of a system of two spin-1/2 particles, drawn to emphasize the analogy to the continuous wave functions plotted in the earlier figures, with larger magnitudes shown as brighter shades and zero shown as black.  The $z$ component of the spin of the first particle is plotted horizontally, and that of the second particle is plotted vertically.\cite{StyerTwoSpinPlots}}
\label{2by2DensityPlots}
\end{figure}

Second, entangled quantum systems are important.  From atoms and molecules to quantum computers, entanglement is central to a large and growing number of real-world applications.\cite{ResourceLetter}  Students will be better prepared to understand these applications if they become reasonably comfortable with entanglement first, in the relatively familiar context of wave functions in one spatial dimension.  When the time comes to make the transition to a system of two spin-1/2 particles, one could emphasize the correspondence using $2\times2$ ``density plots'' as shown in Fig.~\ref{2by2DensityPlots}.  Unfortunately, I do not know of a good visual representation for the wave function of two particles in three spatial dimensions.

Third, entanglement is essential to the quantum measurement process.  Measurement requires interaction, and this interaction entangles the system being measured with the measurement apparatus,\cite{SternGerlach} as Schr\"odinger first emphasized in his famous ``cat paradox'' paper.\cite{SchrodingerCat}  Students are naturally curious about the cat paradox in particular and the measurement controversy more generally.  Although understanding entanglement is not sufficient to resolve the controversy, it is surely necessary.

\appendix*

\section{History of the term ``entanglement''}

That it took roughly 60 years for the term ``entanglement'' to come into common use, after Schr\"odinger introduced it in 1935, is astonishing.  Explaining the long delay is a job for historians of science.\cite{HistoryBooks}  Here I will merely document some of the relevant publications and dates.\cite{disclaimer}

\begin{figure*}[t]
\centering
\includegraphics[width=13cm]{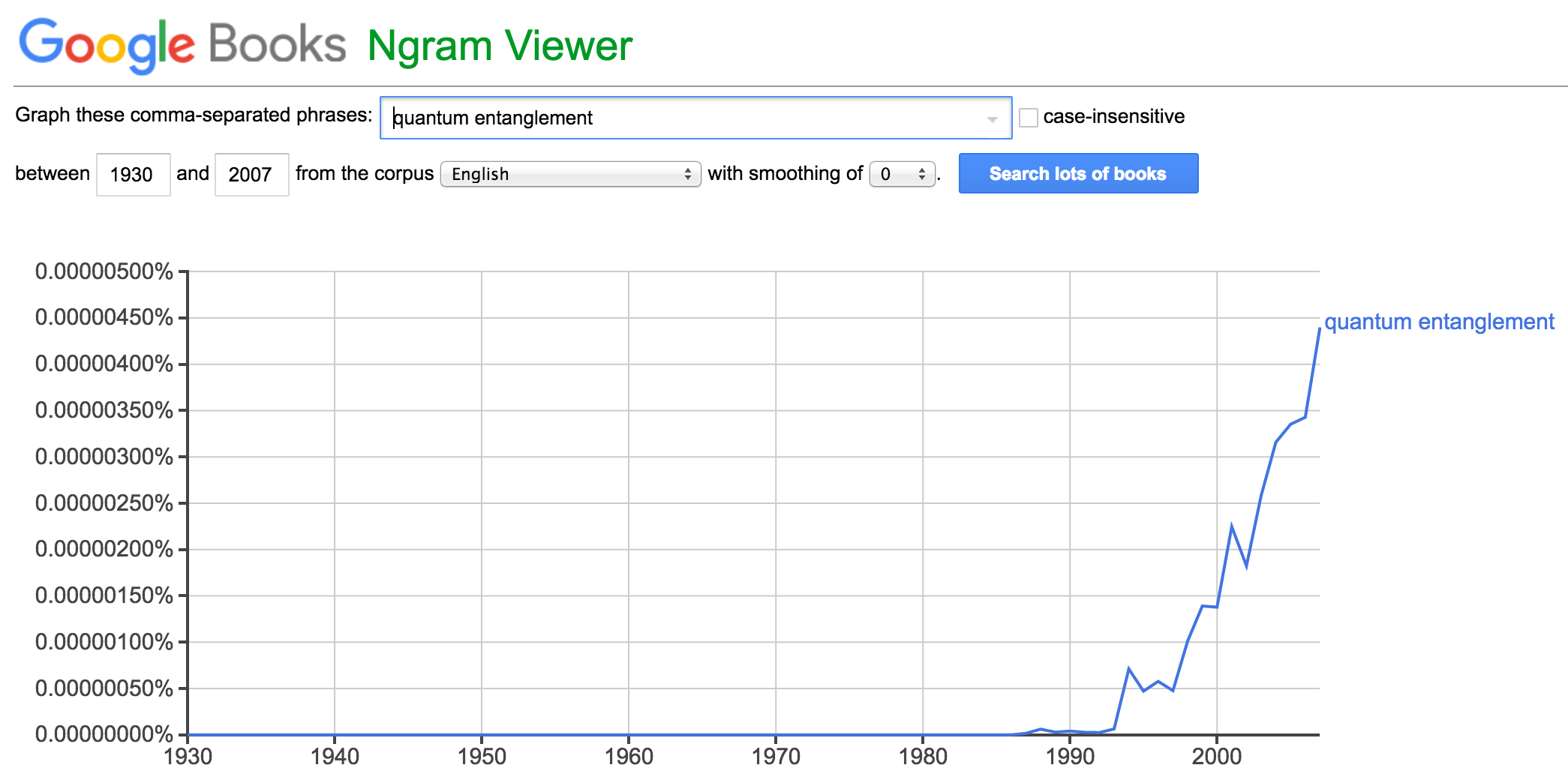}
\caption{Screen capture from the Google Books Ngram Viewer,\cite{Ngram} showing the search results for the phrase ``quantum entanglement.''}
\label{NgramPlot}
\end{figure*}

A convenient way to see the big picture is to use the Google Books Ngram Viewer\cite{Ngram} to search for the phrase ``quantum entanglement.''  As Fig.~\ref{NgramPlot} shows, the phrase does not occur at all in this large database of books until 1987.  There are then a very small number of occurrences through 1993, after which the number rises rapidly.

``Entanglement'' was not completely dormant during the first 50 years after 1935, but its use was sporadic even among specialists in the foundations of quantum mechanics.  The first uses of the term I have found after Schr\"odinger's were by Hilbrand Groenewold in 1946,\cite{Groenewold} by Henry Margenau in 1963,\cite{Margenau1963} and by James Park, a student of Margenau, in 1968\cite{Park1968} (in the American Journal of Physics).  The term continued to appear occasionally in articles on quantum foundations during the 1970s.\cite{1970s}  Then, in a 1980 publication\cite{Bell1980} of a 1979 talk, John Bell pointed out that if we try to explain the ongoing experiments with correlated photons by suggesting that the orientations of the polarizers cannot actually be chosen independently, then ``separate parts of the world become deeply entangled, and our apparent free will is entangled with them.''  An almost identical sentence appears in his better-known ``Bertlmann's socks'' article,\cite{Bell1981} published in 1981.  Peres and Zurek quoted this sentence the following year in the American Journal of Physics.\cite{PeresZurek}

Still, ``entanglement'' remained outside the standard lexicon well into the 1980s.  It does not appear in the 1978 review article on Bell's theorem by Clauser and Shimony,\cite{ClauserShimony} or in the 1984 review article on ``nonseparability'' by d'Espagnat,\cite{dEspagnat} or even in the 1987 resource letter on foundations of quantum mechanics by Ballentine.\cite{Ballentine}  In Wheeler and Zurek's 800-page annotated compilation of papers on quantum measurement theory, published in 1983, ``entanglement'' appears only in the 1935 paper by Schr\"odinger.\cite{SchrodingerCat}  The 1986 book \textit{The Ghost in the Atom},\cite{GhostBook} edited by Davies and Brown, presents interviews with eight experts in quantum foundations (including Bell), none of whom use the word ``entanglement.''

But by 1984, Abner Shimony was saying ``entanglement'' regularly, apparently becoming the term's main champion.  He used the word several times in a paper published that year in a philosophy journal,\cite{Shimony1984} then used it in a Danish television documentary that aired in 1985.\cite{Video1985}  Also in 1985, Nick Herbert's popular book \textit{Quantum Reality}\cite{Herbert} described how two-particle quantum states are not necessarily separable, referring to this property (somewhat confusingly) as ``phase entanglement.''

The decisive year for ``entanglement'' was probably 1987.  A conference was held in London that year to celebrate Schr\"odinger's 100th birthday, and Bell's contribution\cite{Bell1987} to the conference proceedings highlighted Schr\"odinger's phrase ``quantum entanglement,'' even using it as a section title.  That article was also included in the collected volume of Bell's writings on quantum foundations that was published the same year, so it reached many other physicists.  The subject of Bell's article was the newly published (1986) ``spontaneous collapse'' proposal of Ghirardi, Rimini, and Weber (GRW).\cite{GRW1986}  The 1986 GRW paper did not use the word ``entanglement,'' but these authors did use it in 1987 in an answer\cite{GRW1987} to a comment on their paper, and in this answer they cited Bell's contribution to the Schr\"odinger volume.  This answer is the earliest use of the term that I can find in any of the Physical Review journals.

Another year and a half passed before ``entanglement'' appeared in Physical Review Letters, in a paper by Horne, Shimony, and Zeilinger.\cite{HSZ1989}  By then Shimony had also said ``entangled'' once in a Scientific American article,\cite{ShimonySciAm} and used the term repeatedly in his chapter on quantum foundations in \textit{The New Physics},\cite{ShimonyInDavies} a book intended for interested laypersons.  In 1990 he and coauthors Greenberger, Horne, and Zeilinger used it in the American Journal of Physics.\cite{GHSZ}

At that point it was just a matter of time before ``entanglement'' entered the vocabulary of most physicists and interested laypersons.  Roger Penrose used the term in popular books published in 1989 and 1994.\cite{Penrose}  Its first appearance in Physics Today seems to have been in 1991, in Eugen Merzbacher's retiring address as president of the American Physical Society.\cite{MerzbacherPhysToday}  Curiously, in this transcript Merzbacher attributed the term to Margenau, though without any citation.

Getting ``entanglement'' into textbooks took somewhat longer.  The earliest textbook to use the term appears to be Merzbacher's third (1998) edition,\cite{Merzbacher3} which gives a clear and general definition of the concept (and correctly attributes the term to Schr\"odinger).  Five more years went by before it appeared in an undergraduate textbook, Gasiorowicz's third edition.\cite{Gasiorowicz3}  Since then most new quantum mechanics textbooks have mentioned the term at least briefly, although many apply it only to spin systems.

\vfill  

\begin{acknowledgments}

The work of Dan Styer inspired many parts of this paper, and he specifically contributed Fig.~\ref{NgramPlot}.  David Griffiths, Scott Johnson, David Kaiser, David McIntyre, Tom Moore, and Dan Styer read early drafts of the manuscript and provided comments that greatly improved the final version.  I am also grateful to Weber State University for providing a sabbatical leave that facilitated this work.

\end{acknowledgments}


\begin{thebibliography}{99}
\bibitem{CatState}  The term ``cat state'' refers to Schr\"odinger's cat (see Ref.~\onlinecite{SchrodingerCat}), which is supposedly in a superposition of alive and dead states.  Nowadays many physicists use this term to describe any quantum state that is best thought of as a superposition of two other states, especially when the difference between those two states is large or important.  In the example used here the two states are localized around points that are well separated from each other, so in a sense the particle is in two places at once.
\bibitem{IdenticalParticles} Although it is interactions that cause unentangled particles to become entangled, not every entangled state must result from an interaction.  For example, a system of two identical fermions is inherently entangled, although the indistinguishability of the particles and the presence of spin introduce further subtleties.
\bibitem{disclaimer}  This summary is inevitably incomplete, and I would welcome the communication of additions and corrections from readers who are knowledgable about the history of the term ``entanglement.''
\end{thebibliography}
\end{document}